\documentclass[twocolumn,aps,pra,superscriptaddress,nofootinbib,runinaddress,a4paper,showpacs]{revtex4-1}
\usepackage{amsmath}
\usepackage{graphicx}

\newcommand{\be}{\begin{eqnarray}}
\newcommand{\ee}{\end{eqnarray}}
\newcommand{\beq}{\begin{eqnarray}}
\newcommand{\enq}{\end{eqnarray}}
\newcommand{\BEQ}{\begin{eqnarray}}
\newcommand{\ENQ}{\end{eqnarray}}
\newcommand{\EEQ}{\end{eqnarray}}
\newcommand{\forget}[1]{}

\begin{document}
\bibliographystyle{apsrmp}
\title{Can quantum theory and special relativity peacefully coexist?}
\thanks{Invited white paper for Quantum Physics and the Nature of Reality, John Polkinghorne 80th Birthday Conference.  St AnneÕs College, Oxford. 26 - 29 September 2010. Copyright of Oxford University 2010.}
\author{M.P. Seevinck}
\email{m.p.seevinck@science.ru.nl}
\affiliation{Faculty of Science and Faculty of Philosophy, Radboud University Nijmegen, the Netherlands;\\}
\affiliation{Institute for History and Foundations of Science, Utrecht University, the Netherlands.\vskip-0.5cm~}
\date{August 2010}
\maketitle
\forget{
\begin{quote}~\vskip0.01cm\emph{Now it is precisely in cleaning up 
intuitive ideas for mathematics that one is likely to throw out the 
baby with the bathwater.}\vskip0.5em $~~~~~~~~~~~~~~~~~~~~~~~~~~~~~~~~~$J.S. Bell (1990) \citep[p.~106]{bell90}
\\\\
}
\begin{quote}
~\vskip0.01cm\emph{For me then this is the real problem with quantum theory:  the apparently essential conflict between any sharp formulation [of quantum theory] and fundamental relativity. That is to say, we have an apparent incompatibility, at the deepest level, between the two fundamental pillars of contemporary theory \ldots. It may be that a real synthesis of quantum and relativity theories requires not just technical developments but radical conceptual renewal.
}\vskip0.5em $~~~~~~~~~~~~~~~~~~~~~~~~~~~~$J.S. Bell (1986, \citep[p.~9]{bell86})
\end{quote}
~\vskip0.2cm

\section{Introduction}\noindent
This white paper aims to identify an open problem in `Quantum Physics and the Nature of Reality'---namely whether quantum theory and special relativity are formally compatible---, to indicate what the underlying issues are, and put forward ideas about how the problem might be addressed.
\\
\centerline{---\hskip-0.09cm---\hskip-0.09cm---}\\
\noindent
Consider jointly the following two theorems: firstly, the so-called No-Signalling Theorem in quantum theory\footnote{By quantum theory we mean non-relativistic quantum mechanics \emph{\`a la} von Neumann, where the projection postulate need not necessarily be included.}; and, secondly, Bell's Theorem stating that quantum theory is not locally causal\footnote{Importantly, by Bell's theorem we do \emph{not} mean a violation of a Bell-type inequality by quantum mechanical predictions, but \emph{only} a violation of the condition of Local Causality in quantum mechanics. See section \ref{qmlc}.}.  Then, do quantum theory and the theory of (special) relativity indeed ``peacefully coexist'' (A. Shimony, 1984 \citep[p.~227]{Shim}) or is there an ``apparent incompatibility'' here (J.S. Bell, 1984 \citep[p.~172]{bellspeakable})?  If we think the latter is the case---which we will argue one should---, does this ask for a radical revision of our understanding of what (special) relativity in fact enforces? Or are the requirements set by special relativity quite well-understood and should we, therefore, either adopt an un-relativistic approach in any future physics because relativity is false, or, alternatively, find a new formulation of quantum theory that manages to violate Bell's local causality in a relativistically invariant way? Or might it be the case that a conclusive, and well-understood answer to the central question can only be provided by a quantum theory of gravity?

There is no consensus among contemporary philosophers and physicists as to how one should answer these questions.   For example, Albert and Galchen (2009, \cite{albert}) speak of a ``quantum threat to special relativity'', whereas Clark  \emph{et al.} (2010, \cite{clark}) claim that ``[t]he formal compatibility of quantum mechanics with special relativity is highly nontrivial and is in many ways miraculous.'' Is there indeed such a formal compatibility? If indeed so, can we understand this ``miracle''? Or should we agree with Albert and Galchen that we are faced with a severe incompatibility, and in such a way that relativity is likely to be undermined?
 
In this white paper these issues will be addressed as follows.  The doctrine of  ``peaceful coexistence'' between relativity and quantum physics claims simply that relativity and quantum theory will never be found to contradict each other in those cases in which they do happen to be talking about the same things---in particular the types of causal correlations that are possible between events. However, we believe it is fair to claim that no satisfactory proof of this proposition has so far been offered. The reason is the following. Peaceful coexistence is supposed to be guarenteed by the so-called No-Signalling Theorems. However, such theorems in fact presuppose locality of measurements (i.e., a form of no-signalling) and are therefore circular. 

This is a first reason to question the doctrine of peaceful coexistence. But independent of whether one believes this verdict of the no-signalling theorems to be appropriate, it remains the fact that quantum theory is not locally causal and thus violates the causal spacetime structure of relativity. And because of this latter fact a very good case can be made that there really is an incompatibility here between quantum theory and special relativity.  
It thus appears that instead of facing peaceful coexistence we are in fact confronted with armed truce (Peacock, 1991 \cite{peacock}).  

Section \ref{articulating} is devoted to a series of arguments in favour of this position. Section \ref{address} presents a number of ways to address this incompatibility between quantum theory and special relativity. Here we will have to necessarily be brief, both because of lack of space as well as because most of these proposals are still work in progress and not at all clear-cut. Section \ref{conclusion} concludes via a short discussion. Finally, the Appendix  
contains a careful exposition of Bell's notion of local causality.

\section{Articulating the problem and the underlying issues}
\label{articulating}

\subsection{Why peaceful coexistence is not ensured}\label{notpeace}
\noindent
The claim has been made that the principles of quantum theory alone suffice for proofs of the No-Signalling Theorems
\cite{eberhard, GRW, schlieder}.  The most sophisticated proofs of these theorems are in term of quantum field theory and rely on the notion of  \emph{microcausality} or \emph{local commutativity}, which means that operators which represent measurements performed on space-like separate parts of a physical system always commute, regardless of whether or not they would commute if operating locally (Peacock, 1991 \cite[p.~56]{peacock}).

The important observation is that this microcausality condition does not follow from some set of quantum principles, but is in fact postulated because it is ``the mathematical statement of the fact that no signal can be exchanged between two points separated by a spacelike interval and therefore that measurements at such points cannot interfere'' (Schweber, 1961 \citep[p.~723]{schweber}). Indeed, Stapp (1988, \cite[p.~88]{stapp}) admits that ``relativistic quantum field theory \ldots \emph{is constructed} to ensure that its predictions do not depend either on the frame of reference or upon the order in which one imagines performing measurements on spacelike separated regions.'' 

However, ``a proof of a result based on a theory which was `constructed to ensure' that result is no proof at all'' (Peacock, 1991 \citep[p.~70]{peacock}). This conclusion has been endorsed recently by Mittlestaedt (2008, \citep[p.~2]{mittelstaedt}): ``The micro-causality condition of relativistic quantum field theory excludes entanglement induced superluminal signals but this condition is justified by the exclusion of superluminal signals. Hence, we are confronted here with a vicious circle, and the question whether there are superluminal EPR-signals cannot be answered in this way.'' 

Other proofs of the No-Signalling Theorem not using the assumption of microcausality suffer from similar problems because they either implicitly or explicitly assume that measurements have only local effects, thereby begging the question  (Peacock, 1991 \cite{peacock})\footnote{It is noteworthy that Shimony himself abanded his idea of ``peaceful coexistence'', as he admits in (Shimony, 2004 \cite{shimonyStanford}): "The proposal of Òpeaceful coexistenceÓ was in fact espoused at one time by the present author (Shimony, 1978 \cite[section V]{shimony78}), but he was dissuaded from it by a powerful anti-anthropocentric argument of John Bell''. (Shimony here refers to Section 6.12 of (Bell, 1990 \cite{bell90}) where Bell argues that `no-signalling faster than light' cannot be the expression of the fundamental causal structure of contemporary theoretical physics.)}. Of course, what these theorems do show is that the requirement of no-signalling can be worked out consistently in the quantum domain, and as such can be regarded `consistency proofs'. Furthermore, logically, for the desired compatibility between quantum theory and relativity it is not needed that one can derive the no-signalling constraint, or any other relativistic constraint whatsoever.  

But despite this, even if the theorems would be valid, then still the desired compatibility would not be ensured. The reason is that it is highly questionable that special relativity is inextricably bound up with the impossibility of transmitting messages faster than the speed of light.  As for example Maudlin (2002, \cite{maudlin}) has shown, the compatibility of no-signalling and special relativity is much more subtle than this. Special relativity is primarily a theory about the geometrical structure of space and time. And in fact, the truth of the theory is perfectly consistent with theories that have tachyon mechanisms of super-luminal transmission (Maudlin, 2002 \cite{maudlin}; Arntzenius, 1994 \cite{arntzenius}; Berkovitz, 2007 \cite{berkovitz}).

This raises questions about what exactly special relativity enforces, i.e., what the letter of relativity is as opposed to its spirit.
Although we do not want to identify special relativity with the demand for Lorent invariance, we regard it as a particularly clear and uncontroversial part of the theory (Brown, 2005 \cite{brown}). With regard to the possible correlations between outcomes of spacelike measurements in quantum theory, it is fair to claim that, \emph{minimally}, relativity asks for a Lorentz covariant story on Minkowsky spacetime of how the correlations arise. Is this possible? We will see that any attempt to do so and that uses the causal structure implicit in the Minkowskian spacetime faces great difficulty because of Bell's Theorem.

\subsection{Quantum theory is not locally causal: a basic inconsistency with relativity?}
\label{qmlc}\noindent
Bell's condition of local causality is envisaged to encode the Minkowsky spacetime structure for possible physical interactions and influences between physical systems. See Appendix. As Norsen (2007, \cite{norsen07};  2009, \cite{norsen09}) has stressed, particularly noteworthy is the plausibility, generality, and evident appropriateness of Bell's locality criterion as an expression of the relativistic causal structure of Fig.~1 (see Appendix). 
If we now assume that lawlike prediction of correlations is indicative of a causal connection\footnote{Brown (2005 \cite[Appendix~B]{brown}) rejects this view so as to argue that violations of local causality do not entail non-local causation. He remarks: ``Perhaps it is simply not the case that in quantum mechanics, correlations are always apt for causal explanation.'' (Brown, 2005 \cite[Appendix~B]{brown}). We believe this position to be utterly unsatisfactory.}, either directly or via a common cause, then a theory's violation of the criterion of local causality (thereby excluding common causes) means that it posits non-local causation; and not mere non-local correlations. But it should be noted that this not necessarily implies that it supports super-luminal signalling.

As is well known, quantum theory violates local causality, i.e., the theory violates Eq.~\eqref{LC} of the Appendix.  To show this one takes the beable (or `hidden variable') $\lambda$ to be some entangled quantum state $\psi$ (or density matrix $\rho$)\footnote{Indeed, for standard quantum mechanics $\lambda$ is ``already sufficiently specified'' [see second quote by Bell in the Appendix] when the quantum state $\psi$ is fully specified.} and uses suitably chosen observables $a$ and $b$. The formal proof will not be rehearsed here. See for example  Bell (1976 \cite{bell76}, 1990 \cite{bell90}), and many others\footnote{Muller (1999, \cite{muller99}) stresses that no space-time formulation of quantum mechanics is as of yet available---thus it can not be regarded a spacetime theory---, and that it is a hard job to formulate one, be it in Minkovskian or Galilean spacetime. However, despite being true, this is not relevant for the problem here. All that is needed to consider the question of local causality are predictions for measurement outcomes at certain space-time locations as in Fig. 3 (see Appendix), and quantum mechanics \emph{does} give such predictions when the measurements and the state to be measured are specified.  It does not matter that the theory itself cannot be taken to be a spacetime theory on some appropriate differentiable manifold.}.

It is important to comment on some of the facts that are commonly overlooked in obtaining the conclusion that quantum theory violates local causality. Firstly, not needed are Bell's inequalities\footnote{Our conclusion is therefore safe from some common objections against derivations of Bell-type inequalities (e.g., such as problems associated with the need for Kolmogorovian probability theory when averaging over $\lambda$).}.
Secondly, not needed is a `free will' assumption whereby one assumes a form of independence between $\lambda$  and the settings $a,b$.
Thirdly, there is no need for an analysis of the `collapse of the wavefunction' as a real physical process.

It is a rather subtle question---see below---whether or not special relativity genuinely requires
local causality in the sense of Fig.~1 of the Appendix (Maudlin, 2002, \cite{maudlin}). However,  
``if one grants this (and virtually all physicists and commentators do), then it really is possible to establish an ``essential conflict between any sharp formulation [of QM] and fundamental relativity. That is to say, we have an apparent incompatibility, at the deepest
level, between the two fundamental pillars of contemporary theory. . .'' (Bell, 1984 \citep[p.~172]{bellspeakable}).'' (Norsen, 2009 \cite{norsen09}).

\section{Ways to address the problem}\label{address}
\noindent
Can quantum violations of local causality be reconciled with relativity? One can generally distinguish two types of approaches in resolving this issue, and which are not necessarily mutually exclusive:
\begin{enumerate}
\item The approach depends (i) on the interpretation of special relativity (i.e, a specific view on which constraints the theory in fact imposes), or (ii) on some modification of this theory.
\item The approach depends on (i) a specific interpretation of quantum theory, or (ii) on some modification of it.
\end{enumerate}
The approaches of the first type address the problem of reconciling quantum theory with relativity by rejecting that the principle of local causality is implied by relativity; those of the second type that quantum events and/or quantum correlations are of a peculiar nature that just cannot be straightforwardly imbedded into the spacetime structure of Fig.~1 (Appendix).  

In the following we will present a number of approaches of both types, although it will by no means be an exhaustive list:  

{\bf (1)}  A fundamental assumption of the causal structure of Fig.~1 (Appendix) is that measurements (i.e., the settings and outcomes) can be associated with well-defined finite regions in relativistic spacetime. This assumption is needed so that we can assign to each measurement a certain point (or region) in space and at a certain time.
One could choose to reject this assumption. But this has far-reaching consequences: one can then no longer speak of localized events, and it is unclear how one should proceed. Alternatively, perhaps we should adopt a strategy where the wave function ceases to be a function on spacetime and instead becomes a functional on the set of spacelike hypersurfaces? 

{\bf (2)} We need perhaps revise our understanding of what (special) relativity in fact enforces? As Maudlin (2002, \cite{maudlin}) has stressed, we should distinguish between super-luminal signals \emph{simpliciter} and  superluminal signals that allow loops. And only the latter need give rise to inconsistency with relativity. 
It thus seems that the only fundamental relativistic constraint is that of Lorentz covariance (Brown, 2005, \cite{brown}). All talk of super-luminal transmission, signalling, etc. appears to be besides the point.

{\bf (3)} We need perhaps adopt an unrelativistic approach in any future physics because relativity has limited domain of applicability?   Relativity could be only an emergent theory from deeper level physics\footnote{``It seems very likely that relativity, like all other classical theories, will eventually be found to be an approximation to some deeper and (stranger) quantum theory. [\ldots] I find it rather surprising that so many authors have espoused the notion of peaceful coexistence with such confidence. The whole trend in physics in this century seems to rather obviously show that the ultimate breakdown of `peaceful coexistence' is \emph{exactly} what we should expect.'' (Peacock, 1991 \cite[p.~73]{peacock}).}, for example through a holographic scenario (Verlinde, 2010 \cite{verlinde}) or from a noncommutative geometrical theory which is nonlocal with no space and no time in the usual sense, and that are to emerge only in the transition process to the commutative case (Heller and Sasin, 1999 \cite{heller}).

{\bf (4)} Or is it the case that the problem arises because the current model of spacetime as a simple causal manifold is inappropriate? The causal set approach by Sorkin and collaborators (see e.g. (Sorkin, 2010, \cite{sorkin})) seems to be promising in this respect. For them spacetime is a discrete set of spacetime points partially ordered by causal connectibility, which grows by a stochastic process of adding points to the future of the given discrete set; a causal set.

{\bf (5)} Perhaps we can envisage a nonlocal `hidden'-variable model that  'performs the trick' (Gisin, 2009 \cite{gisin_science}) via some non-local influence in spacetime? However, any such attempt is seriously hindered by the fact that any hypothetical nonlocal mechanism that procedures such an influence must be very, very fast (the speed of transmission should at least be four orders (!) of magnitude faster than $c$ (Salart \emph{et al.}, 2008 \cite{salart})). Furthermore, any such a mechanism cannot employ covariant non-local hidden variables $\lambda$ that are invariant under velocity-boosts that changes the time order of events (Gisin, 2010 \cite{gisin2010a}). The only viable option for a non-local mechanism is to assume the existence of a preferred universal frame of reference which univocally determines the time ordering of events. This seems to be a very unwelcome step to take, but it is unclear whether the introduction of a dynamically preferred frame would lead to a gross violation of relativistic causality.  

{\bf (6)} Alternatively, should we find a new formulation of quantum theory that manages to violate Bell's local causality in a relativistically invariant way? Could, for example, Bohmian mechanics be made relativistically invariant? Or can we perhaps obtain some viable relativistically invariant generalisation of non-linear collapse theories? Of the later type, the currently most promising theory is announced by Tumulka (2006, \cite{tumulka1, tumulka2}) who has provided a relativistically invariant dynamical reduction model for many noninteracting fermions. It is interesting in this regard to cite Tumulka (2006, \cite{tumulka1}): ``...we seem to arrive at the following alternative: Bohmian mechanics shows that one can explain quantum mechanics, exactly and completely, if one is willing to pay with using a preferred slicing of spacetime; our model suggests that one should be able to avoid a preferred slicing if one is willing to pay with a certain deviation from quantum mechanics.''

{\bf (7)} Should we adopt a new theory of time and of becoming? 
Gisin (2010, \cite{gisin2010a}) claims that ``quantum events are not merely the realization of usual probability distributions, but must be thought of as true acts of creation (true becoming)'', and in a related paper (2010, \citep[p.~1358]{gisin_science}) he mentions: ``To put the tension [between quantum mechanics and relativity] in other words: no story in spacetime can tell us how nonlocal
correlations happen, hence nonlocal quantum correlations seem to emerge, somehow, from outside spacetime.''\footnote{It is unclear how we should understand the phrase 'happen' here.} He continues (Gisin, 2010 \cite{gisin2010b}): ``Note the implication for the concept of time. Quantum events are not mere functions of variables in space-time, but true creations: time does not merely unfold, true becoming is at work. The accumulation of creative events is the fabric of time.'' Can this indeed be worked out in a full-fledged and understandable theory of time and becoming that resolves our problem?

{\bf (8)} Or can we find solace by merely interpreting quantum theory differently? In the Everett interpretation of quantum theory the threat of non-locality is claimed to be absent\footnote{We believe, however, that this approach is beset with fundamental problems and its solution to the present problem to be far from satisfactory.}: ``[W]hen considering spacelike separated measurements on an entangled system [\ldots] there is no
question of the obtaining of a determinate value for one sub-system requiring that the
distant system acquire the corresponding determinate value, instead of another. Both
sets of anti-correlated values are realised (become definite) relative to different observing
states; there is, as it were, no dash to ensure agreement between the two sides to be a
source of non-locality and potentially give rise to problems with Lorentz covariance.'' (Brown, 2005 \cite[Appendix~B]{brown}).

{\bf (9)} Finally, a serious option to consider is that a conclusive, and well-understood answer to the central question can only be provided by a (yet to be obtained) full-fledged quantum theory of gravity. However, we find this very unlikely, see next section.

\section{Discussion}\label{conclusion}\noindent
The main point of this white paper is to argue that a good and fair case can be made that a basic inconsistency exists between quantum theory and relativity; an inconsistency that is not easily dealt with. Let us hope that  one day the basic inconsistency be illuminated, perhaps harshly, by some new exciting physics. Allow me to end this white paper with the following plea by Norsen (2009, \citep[p.~293]{norsen09}):  
\begin{quote}
``If more physicists would only study Bell's papers instead of relying on dubious
secondary reports, they would, I think, come to appreciate that there really is here
a serious inconsistency to worry about. A much higher-level inconsistency between
quantum theory and (general) relativity has been the impetus, in recent decades, for
enormous efforts spent pursuing (what Bell once referred to as) ``presently fashionable
`string theories' of `everything'.'' (1990, \cite[p.~100]{bell90}) How might a resolution of the
more basic inconsistency identified by Bell shed light on (or radically alter the motivation
and context for) attempts to quantize gravity? We can't possibly know until
(perhaps long after) we face up squarely to BellÕs important insights.''
\end{quote}
What is important here is not so much the appeal to read Bell---although one is strongly advised to do so---but that, instead of focussing on the \emph{higher-level} inconsistency between quantum theory and general relativity, it could very well be a better idea to first resolve the \emph{basic} inconsistency indicated here.   

Thus, as indeed encountered so often, and here once again, the most elementary might not be the easiest to start with.

\acknowledgements\noindent
Useful comments on an earlier version of this white paper have been generously provided by F.A. Muller, Jos Uffink and Ronnie Hermens. 

\noindent
\section*{Appendix: Bell's notion of local causality}
\noindent 
This Appendix\footnote{This Appendix is taken from Seevinck and Uffink (2010, \cite{seevinckuffink2010}), but also see Norsen (2007, \cite{norsen07};  2009, \cite{norsen09}).} outlines Bell's concept of local causality. This is done very carefully because, firstly, there continues to be great misunderstanding among the commentators regarding the status of the concept, and, secondly, so as to convince one of the ``the plausibility, generality, and evident appropriateness of BellÕs locality concept as an expression of the relativistic causal structure \ldots'' (Norsen \cite[p.~15]{norsen07}).
\\
\centerline{---\hskip-0.09cm---\hskip-0.09cm---}\\
\noindent
In the section entitled `Principle of local causality' of the very last article Bell wrote on the foundations of quantum theory  (published in 1990 and entitled 'La Nouvelle Cuisine' \cite{bell90}), Bell begins his explanation of the principle of local causality as defining the causal structure of relativity as follows:\footnote{Here we will mainly focus on Bell's formulation of this principle as presented in 'La Nouvelle Cuisine', Bell (1990) \cite{bell90}. This presentation we take to be the most definite and precise one Bell ever presented; it is overall consistent with earlier formulations Bell used to indicate this principle. See Norsen (2007) \cite{norsen07} for further elaboration and support of this claim.}
\begin{quote} \small
``The direct causes (and effects) of events are near by, and even the indirect causes (and effects) are no further away than permitted by the velocity of light.''  Bell (1990) \cite[p.~105]{bell90}
\end{quote}
~\vskip-0.95cm
\begin{figure}[h]
\includegraphics[scale=0.55]{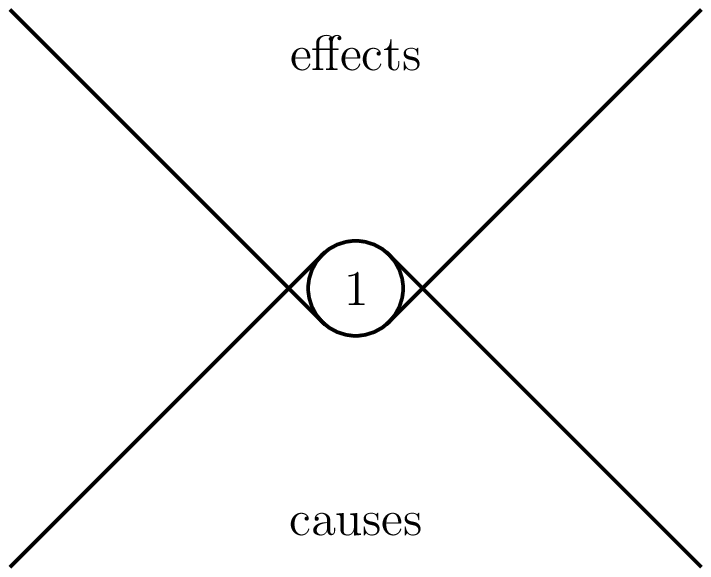}
\caption{``Space-time location of causes and effects of events in region 1.'' Figure (slightly modified) and caption taken from Bell (1990 \cite[p.~105]{bell90}).
} 
\label{figcausal}
\end{figure}
This locates the causes operating in a certain region in spacetime in the backward light cone of that region and effects of anything occuring in that region in its forward light cone. See Fig.~\ref{figcausal}.
 But Bell remarks, ``[t]he above principle is not yet sufficiently sharp and clean for mathematics''. He then continues (see Fig.~\ref{figBell}):
\begin{quote}\small
``A theory is said to be locally causal if the probabilities attached to values of local beables in a space-time region 1 are unaltered by a specification of values of local beables in a space-like separated region 2 when what happens in the backward light cone is already sufficiently specified, for example by a full specification of local beables in a spacetime region 3. It is important that region 3 completely shields off from 1 the overlap of the backward light cones of 1 and 2. And it is important that events 3 be specified completely. Otherwise the traces in region 2 of causes of events in 1 could well supplement whatever else was being used for calculating probabilities about 1. The hypothesis is that any such information about 2 becomes redundant when 3 is specified completely.'' Bell (1990) \cite[p.~106]{bell90}
\end{quote}
\begin{figure}[h]
\includegraphics[scale=0.55]{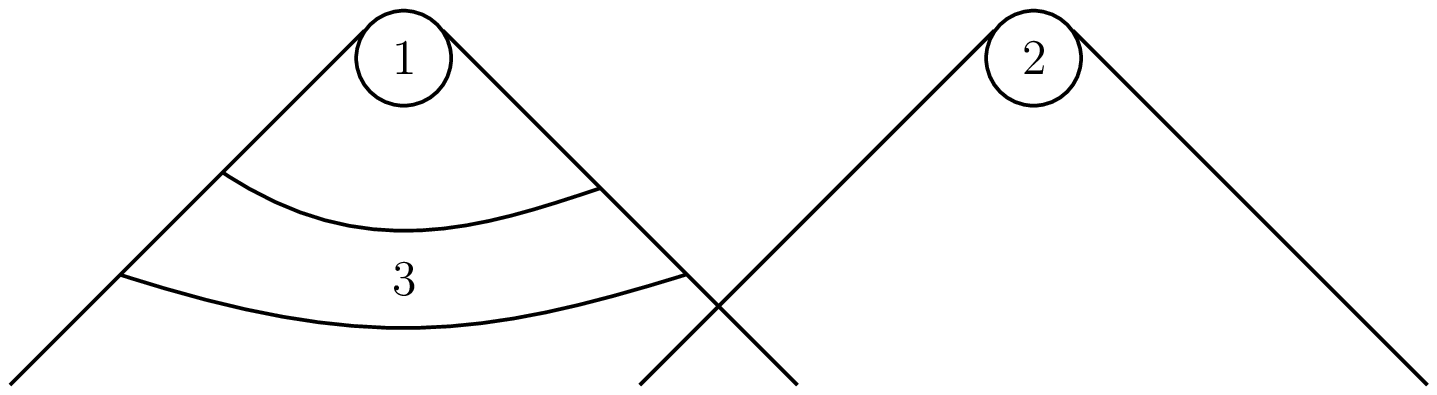}
\caption{``Full specification of what happens in 3 makes events in 2 irrelevant for predictions about 1 in a locally causal theory.'' Figure and caption taken from Bell (1990) \cite[p.~105]{bell90}.} 
\label{figBell}
\end{figure}
Although this formulation is considerably sharper, it is not yet cleanly formulated in terms of mathematics. Probably for this reason Bell introduces some further notation and terminology in a subsequent discussion. 
He in effect introduces the space-time diagram of Fig.~\ref{figBell_fact} that is adapted from Norsen's (2009) \cite{norsen09} highly illuminating paper.
\begin{figure}[h]
\includegraphics[scale=0.6]{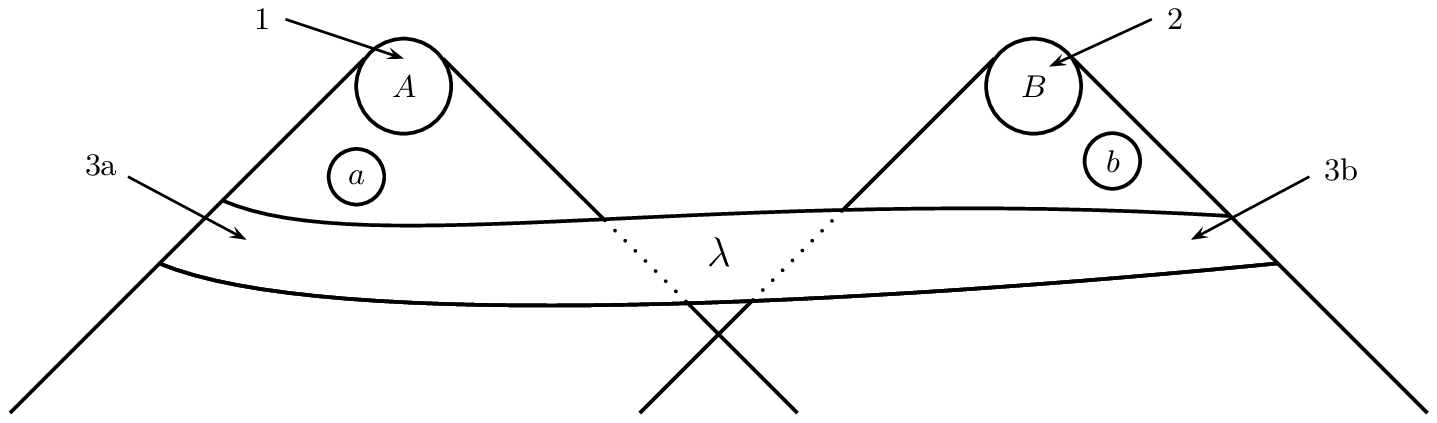}
\caption{Space-time diagram of the setup Bell considers. 
For explanation, see text. Figure adapted from Norsen (2009) \cite{norsen09}.
} 
\label{figBell_fact}
\end{figure}

This diagram encodes the setup Bell considers. It involves measurement on a bi-partite system (e.g., two particles emitted by a source) where each part is measured by a different party, conventionally called Alice and Bob respectively.
The outcomes of measurement are represented by beables $A$ (in region 1) and $B$ (in region 2) and the settings chosen by experimenters Alice and Bob are denoted by beables $a$ and $b$ respectively. The symbol $\lambda$ indicates the specification of the state of the bipartite system under study together with other relevant beables in the spacetime regions 3a and 3b.

The logic is now as follows. Consider a candidate theory that attempts to describe any correlations found between outcomes $A$ and $B$. Suppose region 3a shields off region 1 from the overlap of the past light cones of 1 and 2, and, likewise, that region 3b shields off region 2 from the overlap of the past light cones of 1 and 2 (see Fig.~\ref{figBell_fact}). 
It is assumed that (in this candidate theory under study) $\lambda$ constitutes a complete specification of the beables in region 3a and 3b.

With all this implicitly in place, Bell continues and applies his principle of local causality to this setup:  
\begin{quote}
\small 
``Invoking local causality, and the assumed completeness of \ldots $\lambda$ \ldots we declare redundant certain of the conditional variables in the last expression because they are at space-like separation from the result in question."
Bell (1990)\cite[p.~109]{bell90}\end{quote} 
Thus the  specification of  $\lambda$ makes both $B$ and $b$ redundant for prediction about $A$, and both $A$ and $a$ redundant for prediction about $B$.

This finally allows for a clean formulation in mathematics of the principle. We follow Norsen (2007) \cite{norsen07} in claiming that this indeed gives
\begin{align}\label{LCmath}
\begin{array}{l}
P_{a,b}(A|B,\lambda)=P_a(A|\lambda)\,,
\\
 P_{a,b}(B|A,\lambda)=P_b(B|\lambda)\,,
\end{array}
\end{align}
i.e., the conditional probability\footnote{Note that the settings $a,b$ are indices that label the different conditional probabilities, whereas the outcomes $A,B$ and $\lambda$ are random variables that can be conditioned on. 
See Seevinck and Uffink (2010, \cite{seevinckuffink2010}) for a detailed argument of why this is the appropriate notation, instead of the more common formulation where one also conditions on the settings and thus writes expressions like  $P(A,B|a,b,\lambda)$ and not $P_{a,b}(A,B|\lambda)$ as is done here.} of obtaining $A$ is independent of both $B$ and $b$ given the specification $\lambda$ and $a$, and analogous for the probability of obtaining $B$. Using the definition of conditional probability one trivially obtains the condition
\begin{align}\label{LC}
P_{a,b}(A,B|\lambda)=P_a(A|\lambda)\,P_b(B|\lambda)\,,
\end{align}
i.e., the joint probability for obtaining outcomes $A$ and $B$ factorizes into a product of individual probabilities for the two spatially separated systems, with each factor containing conditionalization only on local beables.  This well-known factorisation condition is thus derived from the principle of local causality, just as Bell himself stressed\footnote{``Very often such factorizability is taken as the starting point of the analysis. Here we have preferred to see it not as the \emph{formulation} of ``local causality'', but as a consequence thereof.'' Bell (1990) \cite[p.~109]{bell90}}.

\end{document}